\documentclass[12pt]{iopart}
\usepackage{graphicx}
\usepackage{amssymb}
\begin{document} 

\title[The number of metastable states in the 
generalized random orthogonal model]{The number of metastable states in the 
generalized random orthogonal model}

\author{R. Cherrier, D.S. Dean and  A. Lef\`evre}
 
\address{IRSAMC, Laboratoire de Physique Quantique, Universit\'e Paul
Sabatier, 118 route de Narbonne, 31062 Toulouse Cedex 04, France}
\ead{lefevre@irsamc.ups-tlse.fr}

\begin{abstract}
We calculate  the number of metastable states in the generalized random
orthogonal model. The results obtained are verified by exact numerical 
enumeration for small systems sizes but taking into account finite size
effects. These results are compared with those for Hopfield model in order
to examine the effect of strict orthonormality of neural network patterns
on the number of metastable states.
\end{abstract}


\section{Introduction}
Mean field or totally connected spin glass models are among the
most widely studied  models of complex systems. They
are the starting point for understanding finite dimensional spin glasses
and are also related to neural network models and complex 
optimization problems \cite{review}. 
Such systems exhibit an exponentially large
number of pure states and dynamic glass like transitions. Below
a certain dynamic transition temperature the dynamics becomes very slow
and the systems stay out of equilibrium on numerical or experimental
time scales. An important factor in the slow dynamics is the 
presence of metastable or blocked configurations. The enumeration
of the number of metastable states has been addressed by various
authors in $p$-spin Ising systems with Gaussian interaction matrices
\cite{tech}, neural network models \cite{gar,tram}, random orthogonal models
(ROMs)\cite{papo}
and in periodic glass models which have no quenched disorder \cite{cis}.
Recently the authors studied the statics of generalized random orthogonal 
models originally introduced in \cite{mapari} 
and established general criteria determining whether these models
exhibit continuous spin glass like transitions or structural glass transitions
preceded by a dynamical transition \cite{cdl}. The first known examples of 
these sorts of phase transitions were seen in $p$-spin models for $p>2$
\cite{pspin}. 
However a number of two spin Ising models were later shown to have 
this structural glass transition \cite{mapari,cdl,deri,dm}. Here we
shall extend the results of \cite{papo} for the ROM to a more general
ROM which can be interpreted as a Hopfield model with strictly orthogonal
patterns in order to explore the influence of pattern orthogonality on the
number of metastable states. We also analyze anti-ferromagnetic Hopfield
model as this  class of  ROMs can also be viewed as 
anti-ferromagnetic Hopfield models. The calculation for the 
standard ferromagnetic Hopfield model was carried out in \cite{gar,tram}
and we also note that the anti-ferromagnetic Hopfield has a structure 
similar to the Hamiltonian arising in the analysis of the Nash equilibria
in the minority game as studied in \cite{dm}. Our analytic results are 
backed up by exact enumeration simulations for small system sizes. 
Despite small system sizes we show that when finite size scaling is taken
into account the agreement with the analytical results is excellent.

The Hamiltonian in a fully connected generalized ROM is
\begin{equation}
H = -{1\over 2} \sum_{ij} J_{ij} S_i S_j
\end{equation}
where the $S_i, \ 1\leq 1 \leq N$ are Ising spins
and
the interaction matrix $J$ is statistically invariant under the 
transformation $J \to {\cal O}^{T} J {\cal O}$ where ${\cal O}\in O(N)$
(the group of orthogonal transformations on $\mathbb{R}^N$).
The matrix $J$ can thus be written as   
\begin{equation}
J = {\cal O}^{T} \Lambda {\cal O}
\end{equation}
where $\Lambda$ is a diagonal matrix with density of eigenvalues
denoted by $\rho(\lambda)$. If the support of $\rho(\lambda)$ is bounded
on the real axis then the thermodynamic limit is well defined.
It was shown in \cite{cdl} that the nature of the spin glass
transition in such models depends on the behavior of the density of states
$\rho(\lambda)$ in the neighborhood of $\lambda_{\rm{max}}$, the largest
eigenvalue of $\Lambda$. 

In this paper we shall concentrate on the model defined by
\begin{equation}
\rho(\lambda) = \alpha \delta(\lambda -1) + (1-\alpha)\delta(\lambda +1)
\end{equation}
The matrix $J$ in this case may be written as
\begin{eqnarray}
J_{ij} &=& \sum_\mu \lambda_\mu \xi^\mu_i \xi^\mu_j \nonumber \\
&=& \sum_{\{\mu\ : \lambda_\mu = 1\}} \xi^\mu_i \xi^\mu_j
- \sum_{\{\mu\ : \lambda_\mu =-1\}} \xi^\mu_i \xi^\mu_j
\end{eqnarray}
where $\xi^{\mu}$ is a random basis of orthonormal vectors on $\mathbb{R}^N$.
One may also write $J$ in the following two forms using the 
completeness of the $\xi^\mu$.
\begin{eqnarray}
J_{ij} &=&  2\sum_{\{\mu\ : \lambda_\mu = 1\}}
\xi^\mu_i \xi^\mu_j  -\delta_{ij}
\label{shift1}\\
&=& \delta_{ij} - 2\sum_{\{\mu\ : \lambda_\mu = -1\}} \xi^\mu_i \xi^\mu_j
\label{shift2}
\end{eqnarray}
We recall that the Hopfield model with $p = \alpha N$
Gaussian patterns has an interaction 
matrix given by
\begin{equation}
J_{ij} = \sum_{\mu = 1}^{\alpha N} \xi^\mu_i \xi^\mu_j
\end{equation}
Here the variables $\xi^\mu_i$ are Gaussian with zero mean 
and variance $\overline{\xi^\mu_i \xi^{\mu'}_j} = 
\delta_{ij} \delta^{\mu \mu'}/N $. These patterns are only orthonormal in the 
statistical sense, that is  $\sum_i\overline{\xi^\mu_i \xi^{\mu'}_i} 
= \delta^{\mu \mu'}$. Up to a constant diagonal term, the 
ROM we study here is from Eq. (\ref{shift1}) equivalent to a ferromagnetic
Hopfield (FH) model 
with $\alpha N$ strictly orthonormal patterns, or from Eq. (\ref{shift2})
equivalent to an anti-ferromagnetic Hopfield (AFH) model 
with $1-\alpha$ patterns.

The number of metastable states gives useful information about the phase 
space of complex systems. The easiest metastable states to analyze are
those which are single spin flip stable, that is to say a configuration 
where flipping a single spin increases (and possibly leaves constant)
the energy of the system. Alternatively every spin is aligned with its local 
field. A metastable state thus defined is a blocked configuration of any 
single spin flip Monte-Carlo dynamics.

\section{Average Number of Metastable States}
In this section following we explain the calculation of the number of 
metastable states for generalized ROMs. By definition the average number of
metastable states is given by 
\begin{equation}
\overline{ N_{MS}} = 
\overline{{\rm Tr}_{S_i} \prod_i \theta(\sum_{j\neq i} J_{ij} S_i S_j)}
\end{equation}
The term $\theta$ is the Heaviside function and is only nonzero if
every spin $S_i$ is aligned with its local field  
$h _i=\sum_{j\neq i} J_{ij} S_j $ {\em i.e.} when $h_i S_i > 0$.
The average number of metastable states at average 
energy $E$ per spin is given by
\begin{equation}
\overline{ N_{MS}(E)} = 
\overline{{\rm Tr}_{S_i} \left[\prod_i \theta(\sum_{j\neq i} J_{ij} S_i S_j)
\right] \delta\left(EN + {1\over 2}\sum_{ij} J_{ij}S_i S_j\right)}
\end{equation}
To proceed we make the gauge transformation 
${\cal O}_{ij} \to {\cal O}_{ij}S_iS_j
= {\cal O}'_{ij}$, it is easy to see that  $ {\cal O}'$ is also in $O(N)$.
One may therefore write  
\begin{equation}
\overline{N_{MS}(E)}  = 
2^N\overline{\left[\prod_i \theta(\sum_{j\neq i} J'_{ij}) \right]
\delta\left(EN + {1\over 2}\sum_{ij} J'_{ij}\right)}
\end{equation}
where $J' = {\cal O'}^T\Lambda {\cal O'}$. Following the standard 
method \cite{tech} we use the identity
\begin{equation}
\theta(x)= \int_0^\infty dx \int_{-\infty}^{\infty} {d\lambda \over 2\pi}\exp(i\lambda x)
\end{equation}
We thus obtain 
\begin{equation}
\overline{N_{MS}} = 2^N \int {d\mu\over 2\pi}{dx_i d\lambda_i\over 2\pi} 
\overline{\exp\left(i\sum_{ij} J_{ij}'(\lambda_i+\frac{\mu}{2}) - i\sum_{i} J'_{ii} \lambda_i -i\sum_{i} \lambda_i x_i + iN\mu E\right)} \label{eqns1}
\end{equation}
To simplify the algebra we make the change of variables
$\lambda_i \to \lambda_i - {\mu\over 2}$.
Following \cite{papo,dgg} we  now consider the term
\begin{equation}
\Omega = \overline{\exp(i\sum_{ij} J_{ij} \lambda_i)} = 
\overline{\exp({i\over 2}{\rm Tr} J(M-2 L))}
\end{equation}
Where in vectorial notation $M = {\bf \lambda} {\bf u}^T +
{\bf u} {\lambda}^T$ with ${\bf u}_i = 1$, 
${\bf \lambda}_i = \lambda_i$ and $L_{ij} = (\lambda_i- \mu/2)\delta_{ij}$.
We note here that in order to eliminate the diagonal term 
in Eq. (\ref{eqns1}) the matrix $L$ has appeared, and this term
must be properly accounted for to obtain the correct result.
The averaging over the $O(N)$ disorder (the Haar measure) in this problem
can be carried out by using the results of \cite{itzu} for $U(N)$ integration and adapting them to the $O(N)$ case \cite{mapari}. Recently a 
simple replica method was also used to derive these results \cite{cdl}.
The results \cite{itzu,mapari,cdl} give for an arbitrary symmetric 
matrix $M$ and $J = {\cal O^{T}} \Lambda {\cal O}$ that
\begin{equation}
\overline{\exp\left[ {1\over 2} {\rm Tr} MJ \right]} =
\exp\left[ {N\over 2} {\rm Tr} G({M\over N}) + {\rm nonextensive\  terms}\right]
\end{equation}
where the overline indicates the Haar average over  $\cal O$ and to leading order the non extensive terms are of order one. A compact 
formula for $G$ is \cite{cdl}
\begin{equation}
G(z) = {\rm max}_{\mu}\{ \mu z - \int d\lambda {\rho(\lambda)\over \mu - 
\lambda} - \ln(z)-1\}
\end{equation} 

In the models of interest here $G(z)$ is given by \cite{cdl}
\begin{eqnarray}
G(z)  &=& {1\over2} \left[\left(1 + 4 z (m + z)\right)^{1\over 2} + m
\ln\left(\left(1 + 4 z (m + z)\right)^{1\over 2} + 2 z + m\right)
\nonumber \right.\\
&-&\left.\ln\left(\left(1 + 4 z (m + z)\right)^{1\over 2} + 1 + 2 mz\right)
- m \ln(m + 1) -1 - \ln(2)\right] \ \ ({\rm ROM}\ m = 2\alpha-1)
\\
G(z)  &=& -\alpha \ln(1-z) \ \ (\hbox{FH})\\
G(z)  &=& \alpha \ln(1+z) \ \ (\hbox{AFH})
\end{eqnarray}
Following the results of \cite{mapari,itzu,cdl}
one obtains
\begin{equation}
\Omega= \exp\left({N\over 2} {\rm Tr}G(i {M-2L\over N})\right)
\end{equation}
Given the form of the matrix $M$, the only non zero eigenvalues of $M$
are in the vector subspace of $\mathbb{R}^N$ spanned by $ {\bf \lambda}$ and ${\bf u}$
\cite{papo}.
The two non zero eigenvalues are
\begin{equation}
\mu_{\pm}
= {\bf \lambda}\cdot{\bf u} \pm |{\bf \lambda}| |{\bf u}|
\end{equation}
We define the order parameters $z$ and $v$ by
\begin{eqnarray}
z &=& {1\over N}\sum_{i} \lambda_i \\
v &=& {1\over N}\sum_{i} \lambda_i^2
\end{eqnarray}
Hence the matrix ${\tilde M} = M/N$ has eigenvalues $z + \sqrt{v}$ and $z -\sqrt{v}$
which are of order $1$ and the other $N-2$ eigenvalues are zero.  We now 
consider the evaluation of the term ${\rm Tr}G(i {M-2L\over N})$, 
Taylor expanding one has
\begin{equation}
{\rm Tr}G(i {M-2L\over N}) = \sum_{n=0}^\infty {i^n\over n!} G^{(n)}(0){\rm Tr}
{(M-2L)^n\over N^n} \label{Tay}
\end{equation}
We note that for finite values of the $\lambda_i$ that to leading order
${\rm Tr} M^p = C(p) N^p$ and that ${\rm Tr} L^p = D(p) N$.
Also for any product $P_p(M,L)$ of the $M$ and $L$ containing $p$ factors 
one has that for $p\ge 1 $  ${\rm Tr} P_p(M,L) < N^{p + \epsilon}$ when $N$ is large for any
small positive $\epsilon$. The dominant terms of this form are 
when $P_p(M,L) = M^p$. To see this,
consider a product $P_p(M,L)$ with at least one 
$L$ occurring, we can thus write, exploiting the Cauchy-Schwartz inequality
\begin{eqnarray}
{\rm Tr} P_p(M,L) &=& {\rm Tr} L P_{p-1}(M,L) \nonumber
\\                &\leq & \left({\rm Tr} L^2\right)^{1\over 2} 
 \left({\rm Tr}P_{p-1}(M,L)^2 \right)^{1\over 2} \nonumber \\
                  &\leq & {\rm Const}\ N^{1\over2} \times N^{p+\epsilon-1} = 
{\rm Const}\ N^{p+\epsilon -{1\over 2}}
\end{eqnarray}
Hence for $p \ge 2$ any product containing at least one $L$ is such that
${\rm Tr} P_p(M,L)/N^p \leq {\rm Const}\ N^{\epsilon-{1\over 2}} \to 0$. 
The only term
containing $L$ which survives the thermodynamic limit is that  in the 
linear term of the Taylor expansion (\ref{Tay}). Putting all this together
for large $N$ yields
\begin{equation}
\Omega= \exp\left(N\left({1\over 2}G(iz + i\sqrt{v}) +{1\over 2} 
G(iz - i\sqrt{v}) - i(z-{\mu\over 2}) G'(0)\right)\right)
\end{equation}
It is easy to show that $G'(0) = \int d\lambda \ \lambda \rho(\lambda)$, 
and hence in the ROM $G'(0) = 2\alpha -1$, which is zero when $\alpha = 1/2$
explaining why the diagonal term mentioned above was unimportant in the 
calculations of \cite{papo,dgg}. It is easy to see heuristically the origin
of this term, as it comes from the term  $D=\sum_{i} J'_{ii} \lambda_i$
in Eq.(\ref{eqns1}), the results of the above calculation is to show that
in the large $N$ limit $D$ can be written as   $D\approx 
({1\over N} \sum_{i} J'_{ii}) (\sum_i \lambda_i ) = 
{1\over N}{\rm Tr}(J)(\sum_i \lambda_i )$.

Introducing a delta function representation for the order parameters, the 
$x_i$ and $\lambda_i$ and $\mu$ integrals may be done yielding
\begin{equation}
\overline{ N_{MS}(E)} = {N^{3\over 2} \over 16\pi^2}
\int dz\ dv\ ds\ dt \left({8\pi\over t}\right)^{1\over 2}
\exp\left(N A[z,v,s,t,E]\right)
\end{equation}
where 
\begin{eqnarray}
 A[z,v,s,t,E] &=& {1\over 2}\left( G(z+i\sqrt{v}) + G(z-i\sqrt{v})\right)
- z G'(0) - sz + {vt\over 2}\\ \nonumber
&+& B({s\over \sqrt{t}})
+ {2\over t}{\left(E + {s\over 2} + {t z\over 2}+ {1\over 2}G'(0)\right)}^2  
\label{actiona3}
\end{eqnarray}
where
\begin{eqnarray}
B(u) &=& \ln\left(\sqrt{{2\over \pi}}\int_{-u}^\infty dx \ \exp(-{x^2\over 2})
\right)
\nonumber \\
&=& \ln\left(1 +  {\rm erf}({u\over \sqrt{2}})\right)
\end{eqnarray}
and anticipating a real action we have made the transformation
$z\to -iz$
The average energy $E^*$ of the metastable states is then given by
\begin{equation}
E^* = -{s\over 2} - {t z\over 2}- {1\over 2}G'(0)
\end{equation}
at the saddle point of the reduced action
\begin{equation}
 A[z,v,s,t] = {1\over 2}\left( G(z+i\sqrt{v}) + G(z-i\sqrt{v})\right)
- z G'(0) - sz + {vt\over 2} + B({s\over \sqrt{t}})\label{actiona}
\end{equation}
The total average number of metastable states is given by
\begin{equation}
\overline{ N_{MS}} = N\int dE \overline{ N_{MS}(E)}
\end{equation}
which gives
\begin{equation}
\overline{ N_{MS}}= {N^{2} \over 8\pi^2}
\int dz\ dv\ ds\ dt 
\exp\left(N A[z,v,s,t]\right)
\end{equation}

We have therefore to leading order
\begin{equation}
S^*= {\ln(\overline{ N_{MS}})\over N} = {\rm {extr}}_{z,v,s,t} A[z,v,s,t]
-{\ln(2)\over N} - {1\over 2} {\ln(\det {\cal H})\over N} +O(1/N)\label{eqnfs} 
\end{equation}
where ${\cal H}$ is the Hessian of $A$ at the saddle point and the term
$O(1/N)$ comes from the nonextensive terms arising in the $O(N)$ disorder
averaging . The fact that the 
leading order correction is $O(1/N)$ means that $S^*$ can be evaluated
by exact enumeration for quite small system sizes when the above finite
size scaling is taken into account.

The extremization of this action
(with four order parameters) seems quite complicated and has a similar
structure to the saddle point encountered in the calculation of the
average number of metastable states in the periodic glass model
studied in \cite{cis} the Hopfield model \cite{tram,gar} and the 
ROM at $\alpha = 1/2$ \cite{papo,dgg}. The saddle point equations giving
$s$ and $t$ are
\begin{eqnarray}
s &=& {1\over 2}G'(z + i\sqrt{v}) + {1\over 2} G'(z - i\sqrt{v}) - G'(0)
\\
t &=& -{i\over 2\sqrt{v}}(G'(z + i\sqrt{v}) - G'(z - i\sqrt{v}))
\end{eqnarray}
The remaining saddle point equations must be solved numerically. Given the 
definition of the order parameter $v$ we look for solutions to the saddle
point equations with positive $v$. We shall see that the solutions we find
with this prescription agree perfectly with the results of exact enumeration
of small systems where we can calculate both $S^*$ the entropy of metastable
states and $E^*$ the average energy of these states.

\begin{figure}[tbp]
\begin{center}
\includegraphics[width=.6\hsize]{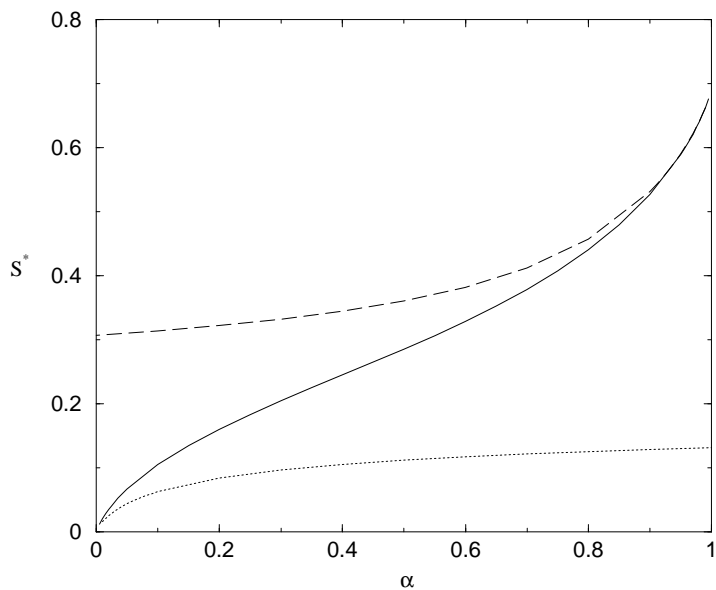}
\end{center}
\caption{\label{figsstar}The entropy of metastable states per spin for
(i) ROM (full line) (ii) FH with $p =\alpha N$ patterns
(dotted line) and (iii) AFH with 
$p= (1-\alpha)N$ patterns (dashed line)}
\end{figure}

The calculated values of $S^*(\alpha)$ for the ROM and ferromagnetic 
Hopfield models
along with $S^*(1-\alpha)$ for the anti-ferromagnetic Hopfield model are
shown in Fig. (\ref{figsstar}). As the ROM can be regarded
as an anti-ferromagnetic Hopfield model with $1-\alpha$ orthonormal patterns
or a ferromagnetic Hopfield model with $\alpha$ orthonormal patterns, this 
comparison is natural. In the limit $\alpha \to 1$ we see in Fig. 
(\ref{figsstar}) that $S^*_{\rm ROM}(\alpha) \to S^*_{\rm AFH}(1-\alpha)$ and
that in these two cases $S^* \to \ln(2)$. However $S^*_{FH} 
\approx 0.131486$.
Hence, for the ROM, as $\alpha \to 1$ 
there is a small fraction $(1-\alpha)$ of repulsive
patterns to be avoided to minimize the energy and the fact that
they are strictly orthonormal or statistically orthonormal does not change the 
behavior of $S^*$ drastically with respect to the AFH. 
This result can be seen analytically as follows.  Writing $\alpha'
= 1-\alpha$, then near $\alpha = 1$, for the ROM
\begin{equation}
G(z) = z - \alpha'\ln(1 + 2z) + O(\alpha'^2)
\end{equation}
In the metastable state calculations here the term linear in $z$ of $G(z)$
disappears from the calculation. The remaining term is just (up to a rescaling
of the energy that will not affect the number of metastable states) the 
term one has for the AFH with $\alpha'$ patterns. Thus explaining
the convergence
of $S^*(\alpha)$ for the ROM with $S^*(1-\alpha)$ for the AFH
near $\alpha =1$. One may further show that in these two cases as $\alpha
\to 1$ one has
\begin{equation}
S^*( \alpha) \approx \ln(2) - {\alpha'\over 2}
\left[\ln\left({2\over e\alpha'} \ln({1\over \alpha'})\right) + {1\over \ln({1\over \alpha'})}\right]
\end{equation} 
This asymptotic formula  agrees well with the numerically calculated
value up to $\alpha' = 0.1$.

In the limit $\alpha \to 0$ we also
see that  $S^*_{\rm ROM}(0^+) = 0$ and  $S^*_{\rm FH}(0^+) = 0$
but  $S^*_{\rm AFH}(1) \approx 0.306983$. However $S^*_{\rm ROM}$ and
$S^*_{\rm FH}$ remain different as $\alpha \to 0$. Hence when the 
attractive patterns are strictly orthonormal then there are more 
metastable states than if the patterns are only statistically orthonormal.   
In the FH model in the limit $\alpha \to 0$ it was shown \cite{gar} that
$S^*(\alpha) \approx {1\over \alpha}[\ln(2/\pi \alpha) -1]$. In the same
limit in the ROM the asymptotic behavior is rather singular and we have
not yet found the corresponding asymptotic behavior.

To summarize we  have the 
inequality $S^*_{\rm FH}(\alpha)\leq S^*_{\rm ROM}(\alpha)
\leq  S^*_{\rm AFH}(1-\alpha)$. Hence in the case of ferromagnetic
Hopfield models strict pattern orthonormality increases the number of
metastable states but in the anti-ferromagnetic Hopfield model 
it decreases the number of metastable states. Let us note here that
at a fixed pattern number the ROM and AFH have more metastable states
than the FH. This is in accordance with the observation that the ROM and
AFH have a structural glass transition whereas the FH has a continuous
spin glass transition.

\section{Numerical Simulations}
To verify our results we have carried out exact numerical
enumeration of $N_{MS}$ on small systems of size between 
$10$ and $30$ spins. Using the finite size scaling predicted
by Eq. ({\ref{eqnfs}) we find excellent agreement between the calculations
presented here even for the relatively small system sizes examined. 

Measured in the simulations were $\overline{N_{MS}}$, $\overline{\ln(N_{MS})}$
and $E^*$ the average value of the energy per spin of the metastable states.
The numerical results confirm to high precision that  $\ln(\overline{N_{MS}})
 = 
\overline{\ln(N_{MS})}$, thus confirming that the total entropy
on metastable states is self averaging and justifying our annealed
calculation. Averaging was carried out over up to $2^{35-N}$ samples for
the systems size of size $N$ , with $N$ between $10$ and $30$. 
Both the annealed
total entropy of metastable states $NS^*=\ln(\overline{N_{MS}})$ and 
the quenched total entropy 
 $N S^*_q\overline{\ln(N_{MS})}$ were plotted as function of $N$ and were found
to be very close to straight lines for systems of size greater than $10$.
The value of $S^*$ was then determined by a linear fit. The average
energy per spin over all metastable states and samples, corresponding to 
the  annealed calculations carried out here, 
\begin{equation}
E^* = {\int dE \ E \overline{N_{MS}(E)}\over\int dE  \overline{N_{MS}(E)}}
\end{equation}
was calculated from the simulations of the  systems of size 20
(to have good statistics). As an 
additional check the quenched average energy per spin of the metastable
states
\begin{equation}
E^*_q = \overline{\left(\int dE \ E{N_{MS}(E)}\over\int dE {N_{MS}(E)}\right)}
\end{equation}
was also calculated. We note  that if annealed approximation is exact then
we should find $E^* = E^*_q$. 

The results for $S^*$ estimated from the annealed average 
 $\ln(\overline{N_{MS}})$ and the quenched average
$\overline{\ln(N_{MS})}$ in the AFH are shown in Fig. (\ref{hfig}) against
the calculated value. We see again that the agreement is excellent. The 
difference between the annealed and quenched averages
are very small showing the validity of the annealed approximation.
Note that by Jensen's inequality the the annealed 
average should be greater than the quenched one. Similarly the 
numerically estimated values for $E^*$ and $E_q^*$ are
within the error bars of the simulation and also in excellent agreement
with the calculated value of $E^*$.

\begin{figure}[tbp]
\begin{center}
\includegraphics[width=.6\hsize]{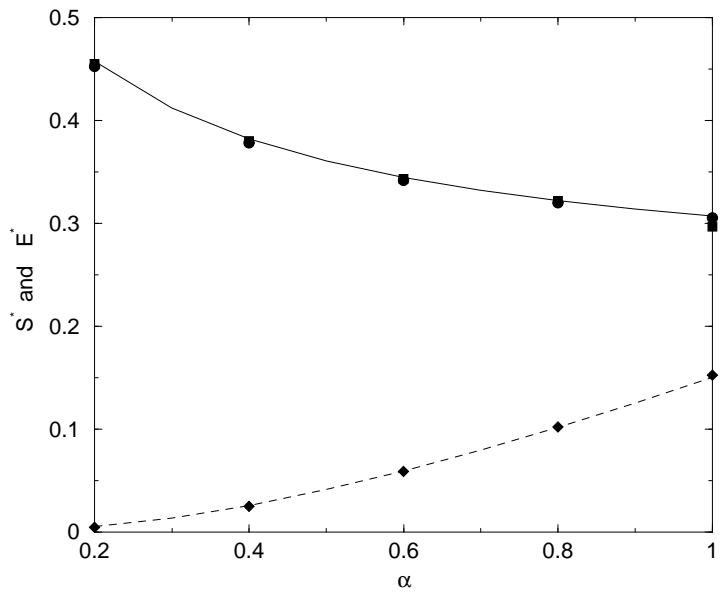}
\end{center}
\caption{\label{hfig}The annealed ($S^*$) (circles) and quenched ($S_q^*$) 
(squares) entropy of 
metastable states for the anti-ferromagnetic Hopfield model measured
from the simulations along with the calculated value (solid line).
Also shown is the average values of the energy per spin of the 
metastable states measured by the simulations (diamonds) along with
the calculated value $E^*$.}
\end{figure}

For the ROM,  in each system  of size $N$, $N\alpha$ eigenvectors 
were chosen to have eigenvalue $1$ and the remaining to have eigenvalue $-1$.
The extrapolated values of $S^*$ and $S^*_q$ are shown in Fig.(\ref{rom.fig})
along with the energy $E^*_q$ obtained from samples of size $N=30$.
We see that again that agreement with the analytical calculations 
is excellent and that the extrapolated values of $S^*$ and $S^*_q$ coincide.

\begin{figure}[tbp]
\begin{center}
\includegraphics[width=.6\hsize]{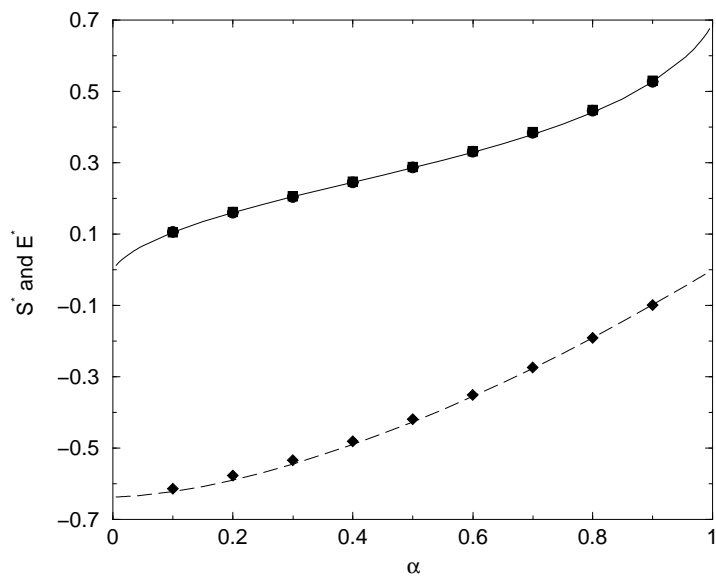}
\end{center}
\caption{\label{rom.fig}As in Fig. (\ref{hfig}), but for the generalized
ROM. For the average energy, the errors where computed and are always 
less than the size of the symbols.}
\end{figure} 

Via the  exact numerical enumeration were also  computed the annealed 
$NS^*(E)=\ln(\overline{N_{MS}(E)})$ and quenched 
$NS^*_q(E)=\overline{\ln(N_{MS}(E))}$ entropy of the metastable states 
of energy $E$. Numerically, we computed
\begin{equation}
\overline{N_{MS}(E_i)}=\frac{1}{\#samples}\sum_{i=1}^{\#samples} 
\frac{N_{MS}(E_i,E_i+\Delta E)}{\Delta E}
\end{equation}
and 
\begin{equation}
\overline{\ln(N_{MS}(E_i))} = \frac{1}{\#samples}\sum_{i=1}^{\#samples} 
\ln\left(\frac{N_{MS}(E_i,E_i+\Delta E)}{\Delta E}\right)
\end{equation}
where $\Delta E = .005$ is the chosen bin size of the discrete energy
values $E_i$.
To avoid divergences when taking the logarithm, we set 
$\ln(N_{MS}(E))=0$ when $N_{MS}(E)$ was zero in a given sample. 
This procedure should be 
unimportant in the thermodynamic limit as it concerns  a 
nonextensive number of metastable states. The results for the ROM 
with $\alpha=1/2$ are  shown in Fig.(\ref{rom2.fig}) for an averaging 
over $32$ samples of size $30$. We see clearly, for a substantial region  
around the most 
probable energy, the annealed and quenched entropies coincide {\em i.e.} 
$S^*(E)=S^*_q(E)$. For values of the entropy smaller 
than $0.3$, the two curves depart from each other, with $S^*_q$ always 
smaller than $S^*$ as it should be. However, 
the small number of metastable 
states considered for these energies and size makes it impossible to draw 
any conclusion about  the thermodynamic limit. In particular one would 
expect that the two entropies should collapse for all energies above some 
energy threshold $E_0$  below $E^*$. This is not the case for our data  
indicating strong finite size effects at the edges of the energy 
spectrum of the metastable states, as one should expect.

\begin{figure}[tbp]
\begin{center}
\includegraphics[width=.6\hsize]{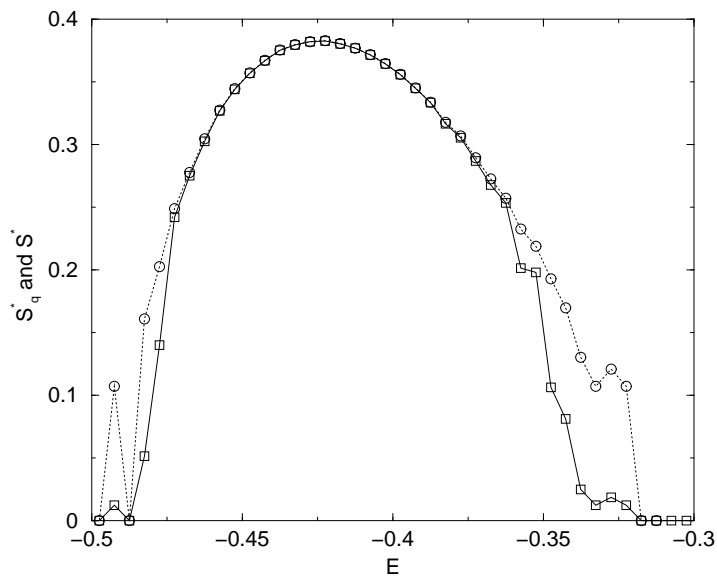}
\end{center}
\caption{\label{rom2.fig}The annealed ($S^*(E)$) (circles, solid line) and 
quenched ($S^*_q(E)$) (squares, dotted line) entropy of metastable states 
of energy $E$ per spin for the ROM.}
\end{figure}

\section{Conclusions}
We have shown that in the class of ROMs considered here that 
there is always an exponentially large number of metastable states
which increases as a function of $\alpha$. The ROM can be viewed
as a ferromagnetic Hopfield model with $\alpha$ strictly orthonormal patterns
or and anti-ferromagnetic Hopfield model with $1-\alpha$ 
strictly orthonormal patterns. Comparison with the corresponding Hopfield
models shows that the orthonormality of the patterns increases the
number of metastable states in the ferromagnetic case but decreases this
number in the anti-ferromagnetic case. If one considers a ferromagnetic 
Hopfield model with all patterns parallel then there are clearly only two 
metastable states (all spins aligned or anti-aligned with this pattern).
In the anti-ferromagnetic Hopfield model if all the patterns are 
parallel then there are more metastable states as it is easier to
be orthogonal to a single pattern than several. This reasoning in 
these extreme cases is compatible with the results found here.

Finally the numerical simulations carried out here, though for 
small system sizes, show remarkable agreement with the analytic
calculations carried out here. This is extremely important as
the structure of the saddle point equations is so complicated
that one needs some confirmation that one has found the good
saddle point. Furthermore it suggests that rather than doing
Monte Carlo simulations for systems exhibiting a dynamical transition,
where even small size systems will stay out of equilibrium, it may be more
useful to carry out exact enumeration on these small
system sizes to calculate thermodynamic quantities. 

\ack During this work we have benefited from
many interesting discussions with A.J. Bray and S.N. Majumdar.

\end{document}